\documentclass[aps,prl,reprint,superscriptaddress,longbibliography]{revtex4-2}

\usepackage[T1]{fontenc}
\usepackage[utf8]{inputenc}
\usepackage{amsmath,amssymb,bm}
\usepackage{graphicx}
\usepackage{xcolor}
\usepackage{hyperref}
\usepackage{orcidlink}

\begin{document}
\title{Apparent Zero-Momentum Signals from Magnon--Magnon Interference in Near-Field Spin-Wave Imaging}

\author{Julien Berthomier}
\affiliation{Laboratoire Albert Fert, CNRS, Thales, Université Paris-Saclay, 1 avenue Augustin Fresnel, 91767 Palaiseau, France}

\author{Romain Lebrun\orcidlink{0000-0002-4109-8388}\email{romain.lebrun@cnrs-thales.fr}}
\affiliation{Laboratoire Albert Fert, CNRS, Thales, Université Paris-Saclay, 1 avenue Augustin Fresnel, 91767 Palaiseau, France}
\author{Vincent Cros\orcidlink{0000-0003-0272-3651}}
\affiliation{Laboratoire Albert Fert, CNRS, Thales, Université Paris-Saclay, 1 avenue Augustin Fresnel, 91767 Palaiseau, France}
\author{Jamal Ben Youssef\orcidlink{0000-0002-9518-9285}}
\affiliation{Lab-STICC, CNRS UMR 6285, Universit\'e de Bretagne Occidentale, 29238, Brest, France}

\author{Karim Bouzehouane\orcidlink{0000-0003-4224-2094}}
\affiliation{Laboratoire Albert Fert, CNRS, Thales, Université Paris-Saclay, 1 avenue Augustin Fresnel, 91767 Palaiseau, France}

\author{Abdelmadjid Anane\orcidlink{0000-0001-5396-6165}\email{madjid.anane@universite-paris-saclay.fr}}
\affiliation{Laboratoire Albert Fert, CNRS, Thales, Université Paris-Saclay, 1 avenue Augustin Fresnel, 91767 Palaiseau, France}

\date{\today}

\begin{abstract}
Spatially resolved measurements of coherent waves are commonly read as direct
maps of the underlying eigenmode spectrum. We show that this interpretation
fails in near-field imaging of multimode spin-wave transport. Scanning
nitrogen-vacancy magnetometry of yttrium iron garnet reveals pronounced
apparent low-wave-vector signals at frequencies for which no propagating
spin-wave mode exists. By tuning the probe--sample distance and modeling the
anisotropic near-field response, we identify their origin as coherent mixing
between directly excited Damon--Eshbach waves and defect- or
transducer-scattered spin waves: the measured signal acquires spatial Fourier
components at the difference wave vector
$\mathbf{k}_{\mathrm{exc}}-\mathbf{k}_{\mathrm{scat}}$, which can vanish even
though both constituent waves carry finite momentum. These results establish
that near-field spectra of coherent excitations are interferometric field
spectra rather than eigenmode maps, providing both a caution and a framework
for identifying scattering processes in nanoscale magnonic devices.
\end{abstract}

\maketitle

The controlled propagation and interference of spin waves offer a route
toward wave-based information processing beyond charge transport
\cite{Pirro2021,Kuznetsov2025,zenbaa_realization_2025, wittrock_soft-x-ray_2026}.
Achieving this objective requires experimental access not only to the
wavelength and amplitude of propagating modes, but also to their phase,
directionality, and scattering into competing channels. Such spectroscopy is
particularly important in nanoscale magnonic devices, where interference,
diffraction, and mode conversion can determine the transmission, phase
stability, and functionality of a circuit
\cite{Guo2026,talmelli_reconfigurable_2020,%
kostylev_spin-wave_2005,stigloher_snells_2016,kiechle_spin-wave_2023}.
A central difficulty, however, is that spatially resolved measurements of
coherent waves do not necessarily provide a direct image of the underlying
eigenmode spectrum.

Scanning nitrogen-vacancy (NV) magnetometry provides a particularly
sensitive platform to expose this physics. A single NV center in diamond
acts as a local, phase-coherent microwave-field sensor, enabling nanoscale
imaging of the stray fields generated by driven magnetic dynamics
\cite{casola_probing_2018,bertelli_magnetic_2020,du_control_2017,%
rondin_magnetometry_2014,finco_imaging_2021}. In contrast with optical
techniques, whose spatial resolution is limited by diffraction, NV
magnetometry accesses spin-wave wavelengths well below the optical
resolution limit while retaining frequency selectivity and phase sensitivity
\cite{sebastian_micro-focused_2015,collet_spin-wave_2017,qin_nanoscale_2021}.
Recent measurements have consequently resolved propagating spin waves,
demonstrating wave-vector filtering by the probe--sample distance and
$\mathbf{k}$-resolution through the occupation of anisotropic isofrequency
contours
\cite{simon_filtering_2022,ogawa_quantitative_2025,manas-valero_isofrequency_2025}.
Yet the observable is the total local microwave field---including that of
the exciting antenna---rather than the population of a single spin-wave
eigenmode. Whenever several coherent contributions coexist, their
interference reshapes the spatial spectrum measured by the NV probe.

Here, we show that this interferometric character produces apparent
reciprocal-space components that do not correspond to any propagating
spin-wave eigenmode. Using scanning NV magnetometry of Damon--Eshbach (DE)
spin waves in a low-damping yttrium iron garnet (YIG) film, we observe
pronounced spectral weight near $\mathbf{k}\approx 0$ at frequencies for
which the spin-wave dispersion forbids propagating modes. By controlling the
NV--sample distance and comparing the measurements with an analytical
near-field diffraction (NFD) model
\cite{temdie_probing_2024,wagle_shaping_2026,vlaminck_spin_2023}, we
identify this signal as a coherent difference-wave-vector contribution
between the directly excited wave and defect- or transducer-scattered spin
waves: the measured contrast contains components at
$\mathbf{k}_{\mathrm{exc}}-\mathbf{k}_{\mathrm{scat}}$, which can approach
zero even though both constituent waves carry finite momentum. This mixing
term governs the measured contrast whenever the spin-wave stray field
dominates over the direct antenna field at the NV position. We identify intrinsic film defects and geometric imperfections of the
microwave transducer as efficient sources of the scattered waves. Beyond clarifying the interpretation of
near-field magnetic images, these results establish coherent momentum mixing
as a ubiquitous feature of low-damping magnonic devices
\cite{dallivy_kelly_inverse_2013,makartsou_spin-wave_2024} and provide a
framework for identifying scattering processes at the nanoscale.

\begin{figure}[t]
    \centering
    \includegraphics[width=0.95\columnwidth]{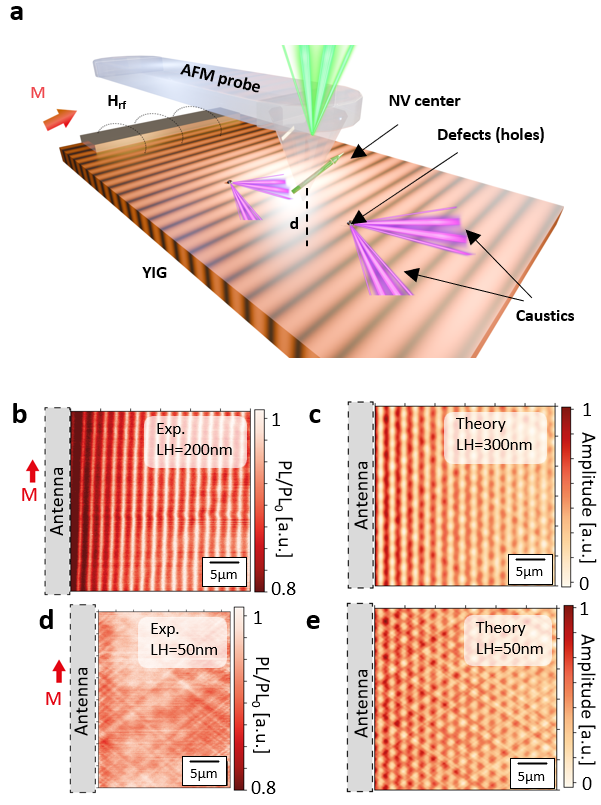}
   \caption{\textbf{Effect of NV-sample distance on phase-coherent spin-wave imaging.} (a) Schematic of the experimental setup: a single NV center at the end of an AFM probe (QZabre) scans above a 140-nm-thick YIG film. A stripline antenna excites primarily Damon--Eshbach (DE) modes with the bias field $\mu_0 H_{\mathrm{ext}} = 15.9$~mT applied parallel to the transducer. Propagating spin waves scatter off defects, generating characteristic caustic beams. (b) Experimental NV contrast at an NV--sample distance $d = 200$~nm, revealing the wavefronts of the coherently excited DE modes. (c) Corresponding near-field diffraction simulation accounting for the interference between the antenna field and the spin-wave stray field. (d) Experimental and (e) simulated magnetic images at a reduced lift height $d = 50$~nm. The emergence of fine-scale interference features demonstrates the increased sensitivity to high-$k$ scattered modes at smaller probe--sample distances. All data were acquired at a drive frequency $f_{\mathrm{NV}} = 2426$~MHz; simulations follow the framework of Ref.~\onlinecite{wagle_shaping_2026}.}
    \label{fig1}
\end{figure}

\begin{figure*}[t]
    \centering
    \includegraphics[width=\linewidth, height=8cm, keepaspectratio]{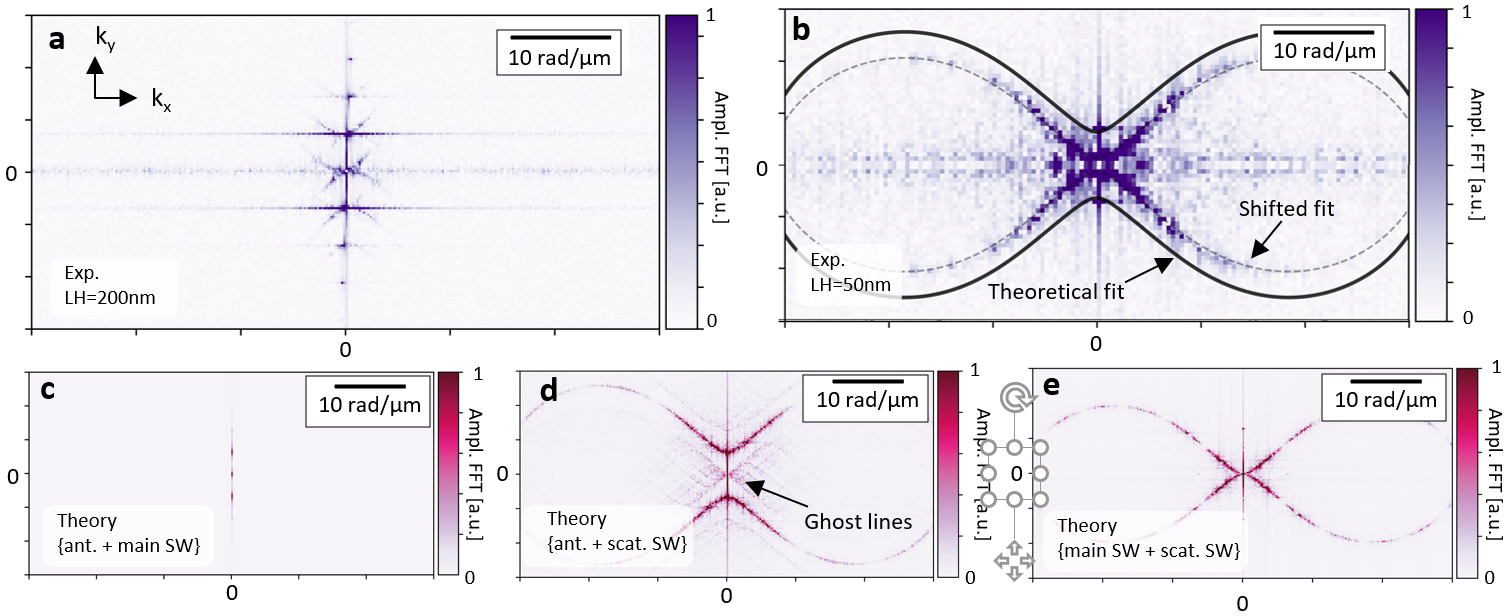}
    \caption{\textbf{Probing the reciprocal space of spin-wave interference at varying lift heights.} (a) Fast Fourier Transform (FFT) of the experimental image at $d = 200$~nm [Fig.~1(b)]. The primary DE modes manifest as distinct reciprocal-space peaks; additional harmonics arise from the nonlinearity of the contrast with respect to microwave power and the spatial anharmonicity of the non-propagating microwave excitation field at the NV position. (b) FFT of the experimental image at $d = 50$~nm [Fig.~1(d)], where a butterfly-shaped spectral distribution emerges with prominent components near $k \approx 0$. The expected analytical isofrequency contour is overlaid as a solid black line. The dashed curve corresponds to the same isofrequency contour but shifted. (c)--(e) NFD simulations of the total microwave magnetic landscape based on Eq.~\eqref{eq:2}: (c) full signal including the antenna and all spin-wave contributions, (d) interference between the antenna field and the scattered modes, and (e) the cross-term between the primary DE packet and the scattered fields, excluding the direct antenna contribution. The latter case accurately reproduces the shifted butterfly contour and the anomalous $k \approx 0$ signal, confirming its origin as magnon--magnon interference. }
    \label{fig2}
\end{figure*}


We investigate spin-wave propagation in a 140-nm-thick YIG film grown by
liquid-phase epitaxy (LPE) \cite{beaulieu_temperature_2018}, with a Gilbert
damping parameter $\alpha = 8\times 10^{-5}$, on top of which a
200-nm-thick, $10$-$\mu$m-wide Au microstrip antenna is patterned. The
antenna is driven by a microwave signal that excites spin waves in the film
while also radiating a direct field in free space. The local magnetic field
is probed by a single NV center embedded in a (100)-oriented diamond
scanning probe [Fig.~\ref{fig1}(a)]. An external magnetic field is applied
along the NV axis and parallel to the stripline antenna, such that spin
waves propagate perpendicular to the magnetization (DE geometry).
The spin-wave signal is accessed through the optically detected magnetic
resonance (ODMR) contrast, $C = 1-\mathrm{PL}/\mathrm{PL}_0$, where
$\mathrm{PL}$ and $\mathrm{PL}_0$ are the NV photoluminescence signals
measured under green-laser illumination ($515$~nm) with and without
microwave excitation. Spatial maps of this contrast reveal interference
fringes associated with propagating spin waves [Figs.~\ref{fig1}(b) and
\ref{fig1}(d)], arising from the interference between the antenna field and
the spin-wave stray field at the NV position
\cite{bertelli_magnetic_2020,simon_filtering_2022,ogawa_quantitative_2025}.
As the probe--sample distance $d$ decreases, the NV becomes sensitive to
shorter-wavelength spin waves, giving access to higher-$k$ components and
revealing scattering-induced interference patterns.
To analyze these patterns, we compute the Fourier transform of the spatial
maps [Figs.~\ref{fig2}(a) and \ref{fig2}(b)]. For NV--sample distances above
$200$~nm, a single propagating mode is observed, the symmetric ($+k$, $-k$)
contribution being an artifact of the Fourier transform of a real-valued
map. This behavior reflects the intrinsic filtering imposed by the
NV--sample distance: at large lift heights, the probe is primarily sensitive
to long-wavelength (small-$k$) modes, since the stray fields of shorter
wavelengths decay evanescently and average out at the NV position. As a
result, only the mode allowed by the DE dispersion at the excitation
frequency is detected.
In contrast, at a reduced lift height of $50$~nm, a butterfly-shaped
spectral distribution emerges in $k$-space, reflecting the two-dimensional
isofrequency contours of spin waves propagating in multiple in-plane
directions. While the high-$k$ features broadly follow the analytical
dispersion \cite{kalinikos_theory_1986} [solid black line,
Fig.~\ref{fig2}(b)], a pronounced and anomalous signal appears near
$k \approx 0$, accompanied by a marked increase in curvature: the observed
contours become significantly steeper as they approach the origin. These
features stand in stark contradiction with the theoretical isofrequency
contour, which predicts a gap in $k_y$ at $k_x=0$ at these frequencies for
propagating modes. Such a discrepancy indicates that the measured signal
does not simply map the intrinsic spin-wave dispersion, but is also shaped
by additional contributions.

To gain deeper insight, we model the spin-wave excitation and the resulting
stray-field distributions within the NFD framework introduced above. This
approach builds on the formalism of Vlaminck \textit{et al.}
\cite{wagle_shaping_2026,vlaminck_spin_2023} and Kalinikos and Slavin
\cite{kalinikos_theory_1986}, which we extend to in-plane magnetization in
order to capture the strongly anisotropic and non-reciprocal dispersion
relations of our experimental geometry (see Supplemental Material
\cite{SM}). Incorporating the specific antenna geometry and magnetic
parameters of the YIG film, the model accurately reproduces the
experimental observations, including the emergence of caustics and the
characteristic dependence of the magnetic contrast on the probe--sample
distance [Figs.~\ref{fig1}(c) and \ref{fig1}(e)].

Crucially, the NFD model allows us to decompose the total microwave field
at the NV position into three distinct physical contributions:
\begin{equation}
\mathbf{B}_{\mathrm{tot}}(\mathbf{r}) = \mathbf{B}_{\mathrm{ant}}
+ \mathbf{B}_{\mathrm{SW,exc}}\, e^{i \mathbf{k}_{\mathrm{exc}}\cdot\mathbf{r}}
+ \mathbf{B}_{\mathrm{SW,scat}}\, e^{i \mathbf{k}_{\mathrm{scat}}\cdot\mathbf{r}},
\label{eq:1}
\end{equation}
where $\mathbf{B}_{\mathrm{ant}}$ is the direct, non-propagating microwave
field radiated by the antenna, $\mathbf{B}_{\mathrm{SW,exc}}$ the stray
field of the primary spin-wave mode directly excited by the antenna, and
$\mathbf{B}_{\mathrm{SW,scat}}$ the stray field of scattered spin waves.
Since the ODMR contrast scales with the local microwave power, the measured
signal follows $S(\mathbf{r}) \propto |\mathbf{B}_{\mathrm{tot}}(\mathbf{r})|^2$;
note that NV imaging thereby gives access to both the amplitude and the
phase of the field. Expanding Eq.~\eqref{eq:1} yields three interference
terms:
\begin{equation}
\begin{aligned}
S(\mathbf{r}) \approx{}& 2B_{\mathrm{ant}} B_{\mathrm{SW,exc}}
\cos(\mathbf{k}_{\mathrm{exc}}\cdot\mathbf{r}) \\
&+ 2B_{\mathrm{ant}} B_{\mathrm{SW,scat}}
\cos(\mathbf{k}_{\mathrm{scat}}\cdot\mathbf{r}) \\
&+ 2B_{\mathrm{SW,exc}} B_{\mathrm{SW,scat}}
\cos\left[(\mathbf{k}_{\mathrm{exc}}-\mathbf{k}_{\mathrm{scat}})\cdot\mathbf{r}\right].
\end{aligned}
\label{eq:2}
\end{equation}
The first two terms describe the conventional interference between the
antenna field and the propagating spin-wave modes, which reproduces the
expected isofrequency contours in $k$-space. The third term, in contrast,
is a magnon--magnon interference between the directly excited and scattered
waves. It introduces a spatial modulation at the difference vector
$\Delta\mathbf{k} = \mathbf{k}_{\mathrm{exc}} - \mathbf{k}_{\mathrm{scat}}$.
Because $\mathbf{k}_{\mathrm{exc}}$ and $\mathbf{k}_{\mathrm{scat}}$
originate, in the linear regime, from the same isofrequency contour, their
difference can approach zero [dashed line in Fig.~\ref{fig2}(b)], providing
a clear physical origin for the anomalous low-$k$ features and the
distorted spectral curvature observed in our measurements.

Importantly, the analytical framework allows a selective decomposition of
these interference terms, providing a direct means to isolate the physical
origin of each feature in the $k$-space maps. Switching off the third term
in Eq.~\eqref{eq:2}, the simulated maps [Fig.~\ref{fig2}(d)] still exhibit
ghost lines in which the spin-wave contours appear shifted toward
$k \approx 0$. These features do not represent physical modes; they are
higher-order Fourier components arising from the inherent anharmonicity of
$S(\mathbf{r})$ and from the nonlinearity of the ODMR contrast with respect to microwave power.
The third term, restored in Fig.~\ref{fig2}(e), introduces components at
$\mathbf{k}_{\mathrm{exc}}-\mathbf{k}_{\mathrm{scat}}$ that can approach
zero even when both contributing modes carry finite wave vectors. It
therefore produces apparent low-$\mathbf{k}$ features that do not
correspond to propagating spin-wave eigenmodes, but instead arise from
spatial beating between coherent finite-wave-vector fields: the measured
reciprocal-space map is distorted relative to the intrinsic spin-wave
dispersion.

Our simulations (see Supplemental Material \cite{SM}) show that this term
dominates when the spin-wave stray field exceeds the antenna field
($B_{\mathrm{SW,scat}} > B_{\mathrm{ant}}$), as occurs in the DE geometry,
where spin waves are efficiently excited and remain localized near the top
film surface. In this regime, the NV response is governed by spin-wave
interference rather than by a direct mapping of the intrinsic dispersion.
The observed low-$k$ features thus constitute an unambiguous signature of
standing spin-wave patterns arising from the coherent superposition of the
main DE mode and the scattered spin waves. These observations, supported by
our calculations, provide direct evidence that NV centers can probe
standing spin-wave patterns at the nanoscale and that such standing waves
radiate a significant microwave field---opening perspectives in which NV
centers are spatially addressed with sub-50-nm resolution and prepared in
specific quantum states.

Conversely, these low-$k$ features are absent, in both experiment and
simulation, in the backward-volume geometry, where the external field is
applied parallel to the propagation direction. In this configuration, spin
waves are excited with significantly lower efficiency and exhibit reduced
surface confinement \cite{bhaskar_backward_2020}, so that the antenna field
remains the dominant contribution to the total microwave landscape and the
signatures of magnon--magnon interference are strongly suppressed. This
confirms that the apparent low-$k$ spectral components emerge specifically
in the regime where the spin-wave stray field is comparable to or exceeds
the antenna field. We note that the mixing term should likewise dominate
for forward- and backward-volume spin waves when using ground--signal--ground
transducers, which generate a negligible direct stray field.
\begin{figure}[t]
    \centering
    \includegraphics[width=0.95\columnwidth]{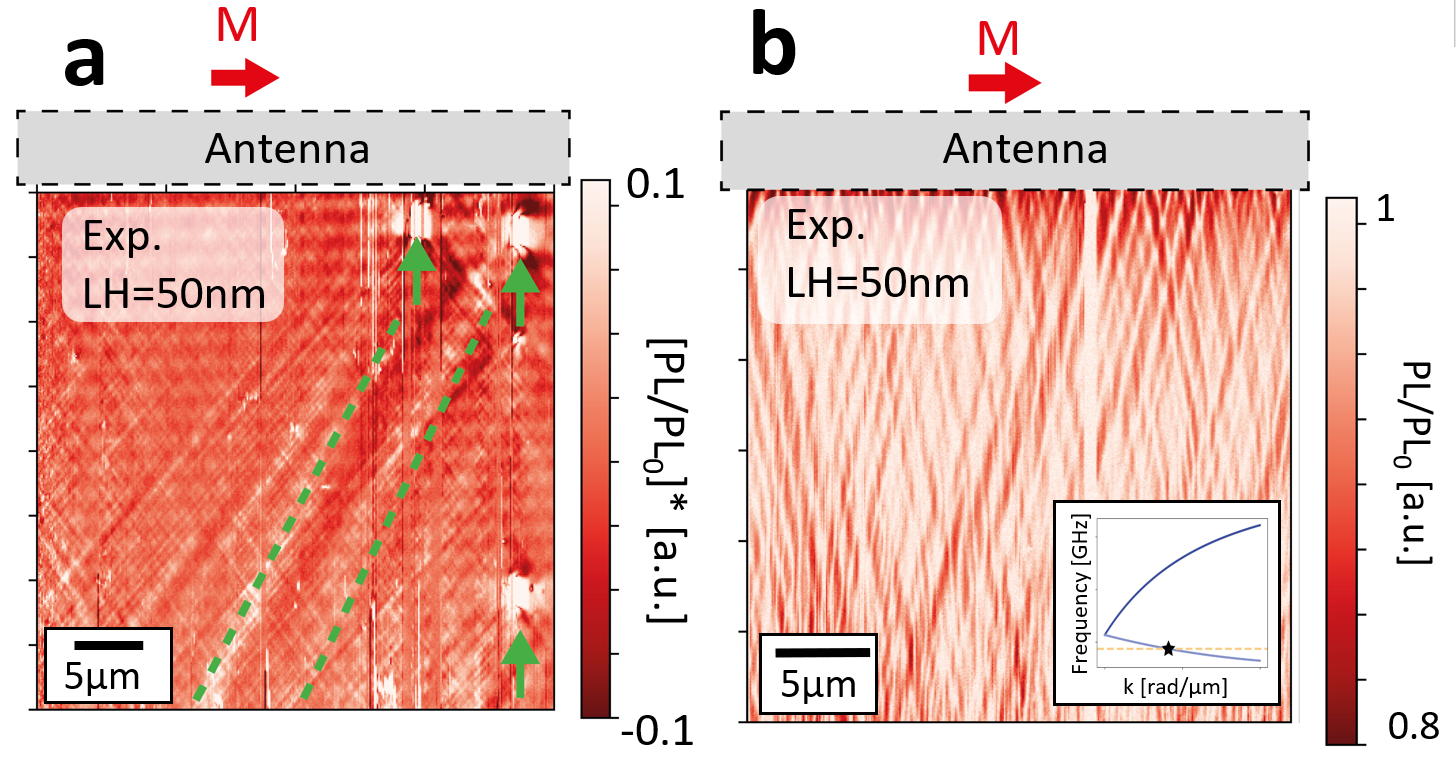}
    \caption{\textbf{Spin-wave scattering from intrinsic defects and transducer corrugations.} (a) Experimental magnetic image revealing caustic beams (dashed lines) originating from scattering centers at the YIG surface (arrows). At $\mu_0 H = 17.1$~mT and $f_{\mathrm{NV}} = 2390$~MHz, the excitation frequency intersects the DE branch, leading to high-efficiency propagation. The contrast $[\mathrm{PL}/\mathrm{PL}_0]^*$ denotes the signal after median-column subtraction to enhance wavefront visibility. (b-inset) Calculated dispersion relation at $\mu_0 H = 33.3$~mT; at $f_{\mathrm{NV}} = 1937$~MHz, the frequency lies below the DE manifold but remains resonant with the backward-volume branch. (b) Experimental NV images at $f_{\mathrm{NV}} = 1937$~MHz. While primary DE modes are energetically forbidden, the presence of scattered wavefronts confirms that transducer corrugations and surface defects act as secondary excitation sources. These features support the near-field diffraction model, where sub-100~nm geometric imperfections seed the interfering modes responsible for the anomalous spectral features.}
    \label{fig3}
\end{figure}

We now turn to the origin of the observed scattering processes. As
suggested in Ref.~\onlinecite{simon_filtering_2022}, defects in YIG can
give rise to caustics \cite{zhou_magnon_2021}. Indeed, holes in the film
resulting from the LPE growth process [arrows in Fig.~\ref{fig3}(a)] act as
point-like defects from which spin waves scatter, producing caustic beams
[dashed lines in Fig.~\ref{fig3}(a)]. Microfabrication-induced corrugations
can, however, also generate scattering patterns, as proposed for wave-based
computing by Papp, Greil and co-workers \cite{papp_nanoscale_2017,Greil2026-do}. To demonstrate
experimentally that antenna-induced inhomogeneities of the excitation field
can produce spin waves with a broad wave-vector distribution, we work in
the DE geometry with the magnetic field chosen such that the excitation
frequency lies below the ferromagnetic resonance frequency, i.e., within
the DE excitation gap, while still crossing the backward-volume dispersion
branch [inset of Fig.~\ref{fig3}(b)]. A perfect antenna could not excite
any propagating spin wave in this configuration. Yet NV imaging reveals
persistent scattered spin waves [Fig.~\ref{fig3}(b)]. Since no primary
spin wave is emitted that could subsequently scatter off YIG defects, the
signal must originate from direct emission by antenna corrugations through
their inhomogeneous field distribution.

\begin{figure}[!b]
    \centering
    \includegraphics[width=0.95\columnwidth]{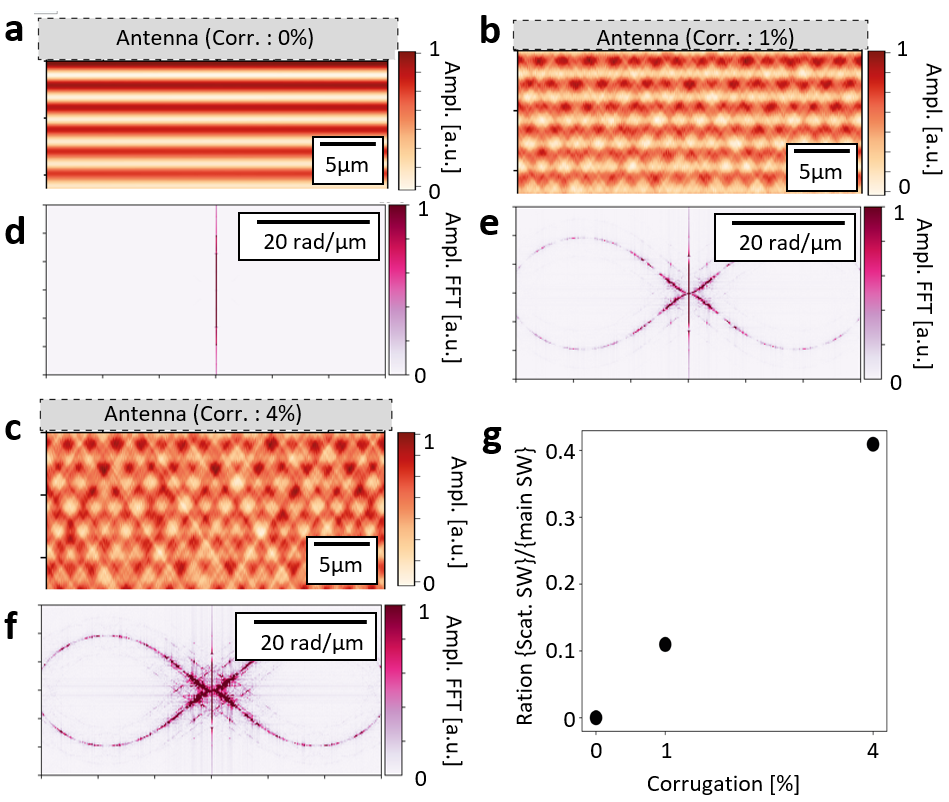}
    \caption{\textbf{Impact of transducer geometry on spin-wave scattering efficiency.} (a)--(c) Real-space NFD simulations of spin-wave propagation excited by antennas with varying edge profiles at a fixed lift height: (a) ideal (rectilinear) edge, where only the primary DE mode is excited, (b) 1\% corrugation width (100~nm), and (c) 4\% corrugations. Insets provide a magnified view of the lithographic transducer definition. (d)--(f) Corresponding spectral distributions in $k$-space obtained via FFT of the real-space profiles. The emergence of high-momentum spectral weight correlates with the presence of edge roughness. (g) Relative amplitude ratio of a representative scattered mode [$\mathbf{k}_{\mathrm{scat}} = (1, 2)~\mathrm{rad/\mu m}$] to the primary excited spin-wave packet as a function of corrugation width. The monotonic increase in this ratio demonstrates that even sub-100~nm geometric imperfections act as coherent secondary sources, significantly shifting the power distribution toward scattered modes in realistic magnonic circuits.}
    \label{fig4}
\end{figure}

To confirm this assertion, we perform NFD simulations in which the
imperfect antenna is modeled by corrugations at its edge, represented as
additional metallic patches of length $1~\mu\mathrm{m}$, randomly
positioned along the edge, and of width $w_{\mathrm{corr}}$. We define the
corrugation percentage as the ratio of $w_{\mathrm{corr}}$ to the
transducer width; e.g., a 4\% corrugation for a $10$-$\mu$m-wide stripline
corresponds to $w_{\mathrm{corr}}=0.4~\mu\mathrm{m}$. The results are
presented in Fig.~\ref{fig4}: the wider the corrugations, the more visible
the scattered spin waves relative to the primary spin-wave packet
[Fig.~\ref{fig4}(g)]. This follows from the fact that less current flows
through narrower corrugations (at constant current density), while their
emitted energy is distributed over higher-$k$ wave vectors that couple
inefficiently to the resonant spin wave at the NV frequency. Nevertheless,
even 1\% corrugations produce visible scattered spin waves, so that they
cannot be neglected in typical lithography processes for magnonic devices.

In conclusion, NV scanning magnetometry emerges as a highly sensitive
technique for imaging spin waves and their scattering processes, with
nanoscale resolution and strong sensitivity to both film imperfections and
lithographic corrugations. This very sensitivity, however, implies that a
proper interpretation of NV measurements requires accounting for the
coherent superposition of the antenna field, the directly excited spin
wave, and the scattered spin-wave fields, whose interference can
significantly distort the measured dispersion. Our results highlight the
critical role of fabrication: edge corrugations as small as 1\% of the
transducer width suffice to generate observable spin-wave interference
patterns, and must therefore be considered both in high-resolution
spin-wave imaging and in the design of standing-wave-based signal-processing
schemes and other coherent magnonic devices.
More generally, our work shows that in realistic systems containing defects
and imperfections, a single pure spin-wave mode cannot be excited: the
energy injected into the magnetic degrees of freedom is inevitably
redistributed among multiple magnonic modes. This multimode population
\cite{devolder_measuring_2021} can lead to insertion losses and phase
noise even in the linear regime, below any nonlinear threshold. A possible
mitigation strategy is to engineer the device geometry so as to lift the
frequency degeneracy of spin-wave modes, thereby reducing unwanted mode
coupling and improving the robustness of magnonic signal propagation.

\begin{acknowledgments}
This work was supported by the French National Research Agency (projects
ANR TACTIQ ANR-23-CE47-0009, ANR SWING ANR-22-EXSP-0004, and
QuanTEdu-France ANR-22-CMAS-0001), the S\'esame project ImageSpin', the
EQUIPEX E-DIAMANT program, and DIM QuanTiP. The authors used AI-assisted
tools for language editing.
\end{acknowledgments}

\bibliographystyle{apsrev4-2}
\bibliography{scattered_spinwaves_refs}

\end{document}


\title{Supplementary Material: Apparent Zero-Momentum Signals from Magnon--Magnon Interference in Near-Field Spin-Wave Imaging}

\author{Julien Berthomier}
\affiliation{Laboratoire Albert Fert, CNRS, Thales, Université Paris-Saclay, 1 avenue Augustin Fresnel, 91767 Palaiseau, France}

\author{Romain Lebrun\orcidlink{0000-0002-4109-8388}}
\affiliation{Laboratoire Albert Fert, CNRS, Thales, Université Paris-Saclay, 1 avenue Augustin Fresnel, 91767 Palaiseau, France}
\author{Vincent Cros\orcidlink{0000-0003-0272-3651}}
\affiliation{Laboratoire Albert Fert, CNRS, Thales, Université Paris-Saclay, 1 avenue Augustin Fresnel, 91767 Palaiseau, France}
\author{Jamal Ben Youssef\orcidlink{0000-0002-9518-9285}}
\affiliation{Lab-STICC, CNRS UMR 6285, Universit\'e de Bretagne Occidentale, 29238, Brest, France}

\author{Karim Bouzehouane\orcidlink{0000-0003-4224-2094}}
\affiliation{Laboratoire Albert Fert, CNRS, Thales, Université Paris-Saclay, 1 avenue Augustin Fresnel, 91767 Palaiseau, France}

\author{Abdelmadjid Anane\orcidlink{0000-0001-5396-6165}}
\affiliation{Laboratoire Albert Fert, CNRS, Thales, Université Paris-Saclay, 1 avenue Augustin Fresnel, 91767 Palaiseau, France}

\maketitle

\section{Near field model}

\begin{figure}[H]
    \centering
    \includegraphics[width=0.45\textwidth]{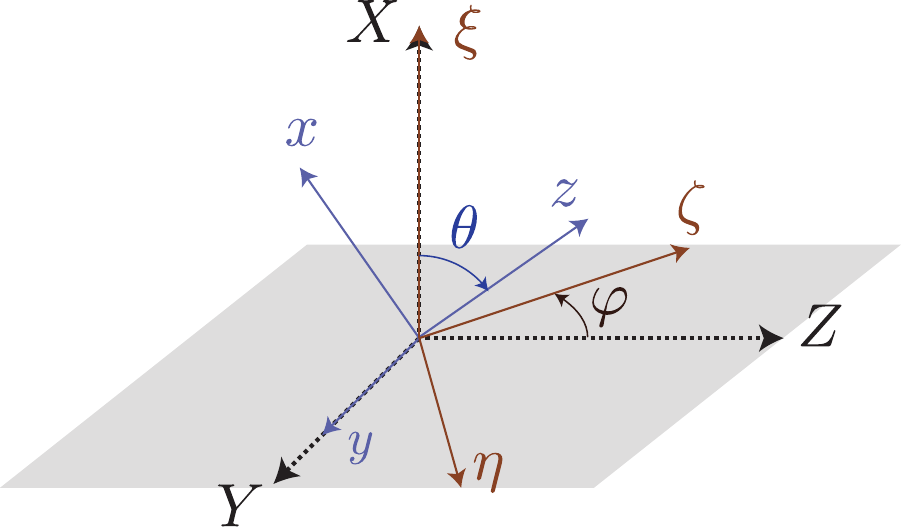}
    \caption{Definition of coordinate systems and rotation angles used in the near-field diffraction framework.}
    \label{fig:angles}
\end{figure}

Simulations in the main text were performed by adapting the near-field diffraction framework of Vlaminck et al. \cite{wagle_shaping_2026}. Deriving this model requires calculating the microwave (MW) field radiated by the antenna alongside the dynamic magnetic susceptibility matrix. The product of these two quantities yields the magnetization vector. From the magnetization vector, and assuming it remains uniform across the sample thickness, we can compute the stray magnetic field at a specific height above the sample surface. To perform these calculations, we define the coordinate systems illustrated in Fig.~\ref{fig:angles}. 

The sample lies in the $(Y,Z)$ plane, such that $X$ defines the out-of-plane axis. The coordinates $(x,y,z)$ describe a rotation of $\theta$ with respect to the coordinate frame $(X,Y,Z)$ so that the static magnetization in the sample is aligned along the local $z$ axis. An in-plane magnetized configuration thus corresponds to $\theta=\frac{\pi}{2}$~rad. The final coordinate system, $(\xi, \eta, \zeta)$, represents a rotation by an angle $\varphi$ of the frame $(X,Y,Z)$ such that a given spin wave propagates in the sample plane along the $\zeta$ axis. Because we restrict our focus to the case where $\theta=\frac{\pi}{2}$~rad, the $y$ axis corresponds directly to $Y$, and the $z$ axis maps to $Z$. 

\subsection{Microwave field from an antenna}

We employ the method described by Bertelli et al. \cite{bertelli_magnetic_2020} to calculate the high-frequency magnetic field generated by a stripline antenna. Given a defined current density profile, the radiated microwave field can be computed within the quasi-magnetostatic approximation using the Biot-Savart law:
\begin{equation}
	\bm{A}(\bm{r}, \omega)=\frac{\mu_0}{4\pi}\int \mathrm{d}^3\bm{r'}\bm{J}(\bm{r'},\omega)\frac{e^{ik\left| \bm{r}-\bm{r'}\right|}}{\left| \bm{r}-\bm{r'}\right|}
\end{equation}

Here, $\bm{A}$ is the magnetic vector potential, $\mu_0$ is the vacuum permeability, $\bm{J}$ is the volume current density, and $k=\frac{\omega}{c}$ is the free-space wavevector of the electromagnetic radiation at angular frequency $\omega$, where $c$ is the speed of light in vacuum. Assuming a uniform current flow through an antenna of thickness $\delta$, the current density profile can be expressed as:
\begin{equation}
	\bm{J}(\bm{r'}=(X',Y',Z'),\omega) = \Theta_{\left[X_i-\delta/2, X_i+\delta/2\right]}(X')\tilde{\bm{j}}(\tilde{\bm{r'}}=(Y',Z'), \omega)
\end{equation}

where $\Theta_{E}(x)=1$ if $x \in E$, and $\Theta_{E}(x)=0$ otherwise, $X_i$ indicates the center coordinate of the antenna along the $X$ axis, and $\tilde{\bm{j}}$ represents the current density distribution within the $(Y,Z)$ plane. Combining these definitions and applying the Weyl identity:
\begin{equation}
	\frac{e^{ik\sqrt{(X-X')^2+(Y-Y')^2+(Z-Z')^2}}}{\sqrt{(X-X')^2+(Y-Y')^2+(Z-Z')^2}}
	= \frac{i}{2\pi}\int \mathrm{d}k_y \mathrm{d}k_z \frac{e^{ik_x \left| X-X'\right|+ik_y(Y-Y')+ik_z(Z-Z')}}{k_x}
\end{equation}

yields the following real-space expression for the potential, valid for $X \leqslant 0$ with $X_i = \delta/2$:
 \begin{equation}
 \bm{A}(X;k_y,k_z, \omega) = i\frac{\mu_0}{2} e^{-ik_x X} \frac{e^{ik_x\delta}-1}{ik_x^2}\mathcal{F}\left\lbrace \tilde{\bm{j}}(\cdot, \cdot, \omega) \right\rbrace (k_x,k_y)
 \end{equation}
 
Operating frequencies in the GHz regime yield a free-space wavevector $k_0 \sim 10^1$~rad/m, which is entirely negligible compared to typical exchange and dipolar wavevectors of excited spin waves ($k_{\mathrm{SW}} \sim 10^6$~rad/m). Given that $k_{\mathrm{SW}} = \sqrt{k_y^2+k_z^2} \gg \omega/c = k_0$, we can safely write:
 $$ k_x = i\sqrt{k_y^2+k_z^2} = i\kappa$$

Using the relation $\bm{B}(\bm{r}, \omega) = \bm{\nabla} \times \bm{A}(\bm{r}, \omega)$, we obtain the components of the driving field in Fourier space:
\begin{equation}
H_Y(X;k_y,k_z) = \frac{1}{2}e^{\kappa X}\frac{e^{-\kappa\delta}-1}{\kappa}\mathcal{F}\left\lbrace \tilde{\bm{j}}\cdot \bm{z}(\cdot, \cdot, \omega) \right\rbrace (k_x,k_y) 
\end{equation}
\begin{equation}
H_Z(X;k_y,k_z) = -\frac{1}{2}e^{\kappa X}\frac{e^{-\kappa\delta}-1}{\kappa}\mathcal{F}\left\lbrace \tilde{\bm{j}}\cdot \bm{y}(\cdot, \cdot, \omega) \right\rbrace (k_x,k_y) 
\end{equation}
\begin{equation}
H_X(X;k_y,k_z) = -i\frac{k_y}{\kappa}H_Y(X;k_y,k_z)-i\frac{k_z}{\kappa}H_Z(X;k_y,k_z)
\end{equation}
In the general case, the driving fields $h_x$ and $h_y$ are recovered by applying the standard rotation matrix of angle $\theta$ from the $(X,Y,Z)$ coordinate system to $(x,y,z)$. For our specific experimental situation where $\theta = \frac{\pi}{2}$~rad, this simplifies to $h_x=H_X$ and $h_y=H_Y$.  

\subsection{Susceptibility matrix}

To model the dynamic magnetization response under in-plane bias, we follow the approach developed by Vlaminck \cite{vlaminck_spin_2023}. Evaluating Kalinikos and Slavin's formulation \cite{kalinikos_theory_1986} to first order in perturbation theory for the fundamental uniform thickness mode in an infinitely thin film yields the matrix equation:
\begin{equation}
\begin{pmatrix}
\frac{\omega_H}{\omega_M}+L_{\text{ex}}k^2 + \sin^2 \theta + \left( \cos^2 \varphi - \sin^2 \theta  \left( 1 + \cos^2 \varphi \right) \right) P & -i\frac{\omega}{\omega_M}+ \sin \varphi \cos \varphi \cos \theta P\\
i\frac{\omega}{\omega_M}+ \sin \varphi \cos \varphi \cos \theta P & \frac{\omega_H}{\omega_M}+L_{\text{ex}}k^2 + \sin^2 \varphi P
\end{pmatrix}\begin{pmatrix}
m_{x,\bm{k}} \\ m_{y,\bm{k}}
\end{pmatrix} = \begin{pmatrix}
h_{x,\bm{k}} \\ h_{y,\bm{k}}
\end{pmatrix}
\end{equation}

Where the physical parameters are defined as:
\begin{itemize}
    \item $\omega_H = \gamma H_i$, where $\gamma$ is the gyromagnetic ratio and $H_i$ is the internal magnetic field.
    \item $\omega_M = \gamma M_0$, where $M_0$ is the saturation magnetization. 
    \item $\theta$ and $\varphi$ represent the out-of-plane angle relative to the sample normal and the in-plane angle relative to the bias magnetic field, respectively.
    \item $L_{\text{ex}} = \sqrt{\frac{2A}{\mu_0 M_0^2}}$ is the exchange length, where $A$ is the exchange stiffness constant. 
    \item $P = \frac{1-\exp(-kL)}{kL}$ is the thin-film geometric reduction factor arising from dipolar boundary conditions, where $L$ is the film thickness. 
\end{itemize}

Gilbert damping is integrated by partitioning the total magnetization into a static background component $\bm{M_0}$ aligned along the local $z$-axis (parallel to $H_i$) and a small dynamic transverse component $\bm{m}$, such that $\bm{M}(\bm{r},t) = \bm{M_0} + \bm{m}(\bm{r},t)$. Taking the temporal Fourier transform, the loss term scales as:
\begin{align}
    \frac{\alpha_G}{M_0} \left( \bm{M}\times \frac{\partial \bm{M}}{\partial t} \right) &= -i\omega \frac{\alpha_G}{M_0} (\bm{M_0}\times \bm{m})\\
    &= -i\omega \alpha_G (\bm{z}\times \bm{m})\\
    &= -i\omega \alpha_G \begin{pmatrix}
        0 & -1 \\
        1 & 0
    \end{pmatrix} \begin{pmatrix}
         m_x \\
         m_y 
    \end{pmatrix}
\end{align} 

We cast the dynamic linear response in terms of the standard volumetric susceptibility tensor:
\begin{equation}
    \bm{m} = \overline{\overline{\chi}}(\omega,\bm{k})\bm{h}
\end{equation}

Inverting the driving matrix directly yields the components of the $2\times 2$ tensor $\overline{\overline{\chi}}$ at a given coordinate $(\omega, \bm{k})$:
\begin{align*}
    \chi_{xx} &= \frac{\omega_H}{\omega_M}+L_{\text{ex}}k^2 + \sin^2 \varphi P + i \alpha_G \frac{\omega}{\omega_M} \\
    \chi_{xy} &= i\frac{\omega}{\omega_M}- \sin \varphi \cos \varphi \cos \theta P \\
    \chi_{yx} &= -i\frac{\omega}{\omega_M}- \sin \varphi \cos \varphi \cos \theta P \\
    \chi_{yy} &= \frac{\omega_H}{\omega_M}+L_{\text{ex}}k^2 + \sin^2 \theta + \left( \cos^2 \varphi - \sin^2 \theta  \left( 1 + \cos^2 \varphi \right) \right) P + i\alpha_G \frac{\omega}{\omega_M} \\
    \chi_{\text{inv}} &= \chi_{yy}\chi_{xx}-\chi_{xy}\chi_{yx}
\end{align*}
yielding:
 \begin{equation}
     \overline{\overline{\chi}} = \frac{1}{\chi_{\text{inv}}}\begin{pmatrix}
        \chi_{xx}  & \chi_{xy} \\
        \chi_{yx}  & \chi_{yy}
     \end{pmatrix}
 \end{equation}

Numerically, the reciprocal space grid is configured to match the fields of view and sampling steps used in real space. For each node in the mesh, the orientation angle $\varphi$ and the local tensor components are evaluated, providing a full map of $(m_x, m_y)$ in Fourier space. An inverse Fourier transform then yields the spatial profiles, which are rotated back into the primary laboratory frame $(X,Y,Z)$.

\subsection{Magnetic field from magnetization}

Once the spatial profile of the dynamic magnetization vector is resolved, the resulting stray magnetic field is calculated following the method of Bertelli et al. \cite{bertelli_magnetic_2020}. Working within the magneto-quasi-static regime ($\bm{\nabla} \times \bm{H} = \bm{0}$), we introduce the magnetic scalar potential $\Phi_M$ defined via $\bm{H} = -\bm{\nabla}\Phi_M$. In the absence of pinning or boundary boundaries at infinity, the potential satisfies:
\begin{equation}
\Phi_M (\bm{r}) = -\frac{1}{4\pi} \bm{\nabla} \cdot \int \mathrm{d}^3 \bm{r'}\frac{ \bm{M}(\bm{r'})}{\left| \bm{r}-\bm{r'} \right| }
\end{equation}
The induction field components can be calculated via:
 \begin{equation}
B_{\beta}(\bm{r},t) = \frac{\mu_0}{4\pi} \partial_{\beta} \partial_{\alpha} \int \mathrm{d}^3 \bm{r'}\frac{ M_{\alpha}(\bm{r'},t)}{\left| \bm{r}-\bm{r'} \right| } 
\end{equation}
where index variables $\alpha, \beta$ select coordinates along the $X,Y$, or $Z$ directions. Projecting this transformation into Fourier space provides algebraic relations for the fields at a vertical position $X$ above the film surface:
\begin{align}
B_X(X;k_Y,k_Z,t)&=\frac{\mu_0}{2}e^{-kX-i\omega t}\left(1-e^{-kL} \right) \left[ M_X(\bm{k})-i\frac{k_Y}{k}M_Y(\bm{k})-i\frac{k_Z}{k}M_Z(\bm{k})\right] \\
B_Y(X;k_Y,k_Z,t)&=-i\frac{k_Y}{k}B_X(X;k_Y,k_Z,t) \\
B_Z(X;k_Y,k_Z,t)&=-i\frac{k_Z}{k}B_X(X;k_Y,k_Z,t)
\end{align}

Applying a two-dimensional inverse Fourier transform yields the quantitative dynamic field vectors mapped over real space coordinates, which are directly evaluated at the position of the NV defect.

\section{Simulation of corrugation}

\begin{figure}[H]
     \centering
     \begin{minipage}{0.48\textwidth}
         \centering
         \includegraphics[width=\textwidth]{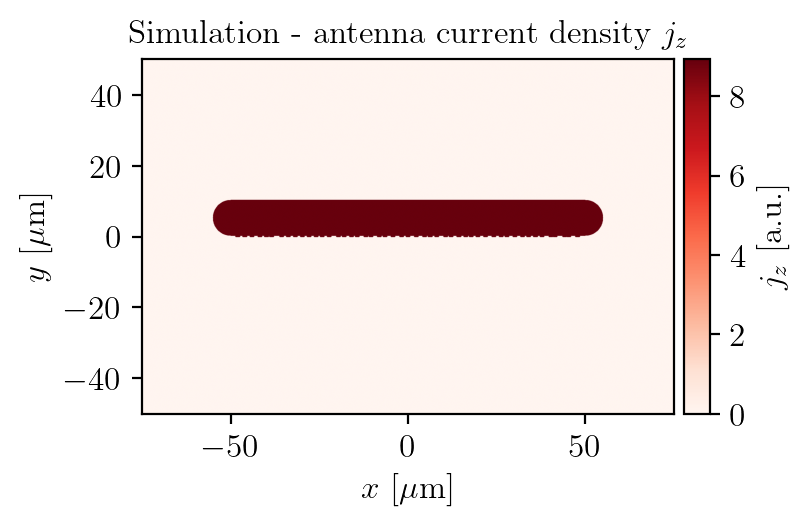}
     \end{minipage}
     \hfill
     \begin{minipage}{0.48\textwidth}
         \centering
         \includegraphics[width=\textwidth]{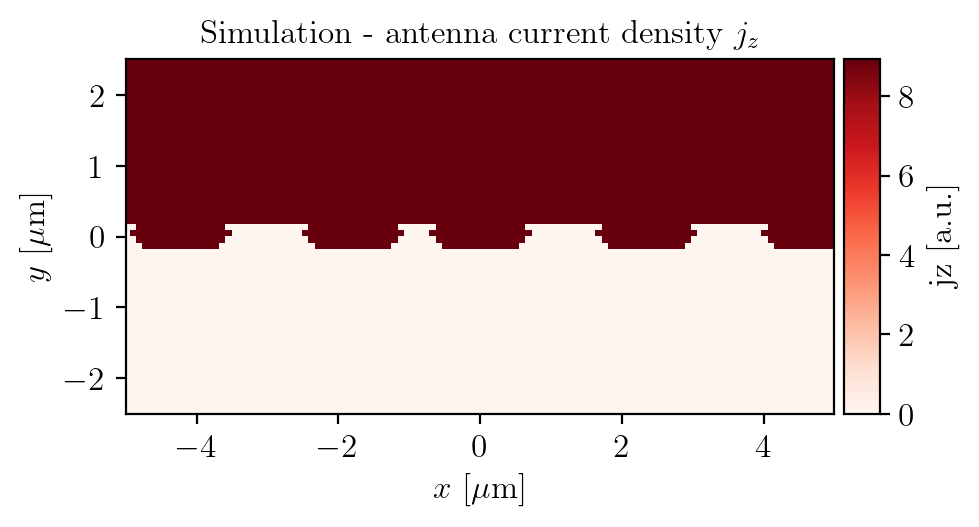}
     \end{minipage}
        \caption{High-magnification view of the simulated current density distribution $J_Z$ for an idealized antenna stripe lacking structural defects (left) and an antenna structured with a 4\% local corrugation ratio (right). The current density profile within the corrugation features matches the magnitude of the primary transmission path.}
        \label{fig:corruguations}
\end{figure}

To simulate the effects of lithographic corrugation on the propagation profiles, we model micron-wide auxiliary metal tracks positioned adjacent to the main driving antenna, assigned a matching peak current density (Fig.~\ref{fig:corruguations}). These localized irregularities source secondary microwave fields that combine coherently with the primary driving profile. The geometric confinement of these corrugations provides high-spatial-frequency components that allow the efficient excitation of a broad continuum of wavevectors $\bm{k}$ in all directions, yielding the distinct multi-mode interference patterns discussed in Fig.~4 of the main text.

\bibliography{scattered_spinwaves_refs}
\label{LastBibItem}